\documentclass[9pt,conference]{IEEEtran}
\IEEEoverridecommandlockouts
\usepackage{cite}
\usepackage{amsmath,amssymb,amsfonts}
\usepackage{bbm}
\usepackage{algorithm}
\usepackage{algpseudocode} % from algorithmicx
\usepackage{graphicx}
\usepackage{textcomp}
\usepackage{xcolor}
\usepackage{subcaption}
\usepackage[normalem]{ulem} 

\def\BibTeX{{\rm B\kern-.05em{\sc i\kern-.025em b}\kern-.08em
    T\kern-.1667em\lower.7ex\hbox{E}\kern-.125emX}}

\newcommand{\va}{\mathbf{a}}

\newcommand{\vx}{\mathbf{x}}

\newcommand{\vs}{\mathbf{s}}

\newcommand{\mY}{\mathbf{Y}}

\newcommand{\mS}{\mathbf{S}}
\newcommand{\mV}{\mathbf{V}}
\newcommand{\mQ}{\mathbf{Q}}

\newcommand{\mX}{\mathbf{X}}
\newcommand{\mA}{\mathbf{A}}
\newcommand{\mW}{\mathbf{W}}
\newcommand{\vw}{\mathbf{w}}
\newcommand{\vy}{\mathbf{y}}

\newcommand{\mD}{\mathbf{D}}

\begin{document}

\title{FastICA with Learned Scores from the Empirical Characteristic Function}

\author{
\IEEEauthorblockN{\textit{David Watts, Jonathan H. Manton}} \\
\IEEEauthorblockA{Dept. of Electrical and Electronic Engineering, The University of Melbourne, Australia \\
dlwatts@student.unimelb.edu.au, j.manton@ieee.org}
}

\maketitle

\begin{abstract}
Independent component analysis (ICA) estimates a demixing matrix that can recover statistically independent sources from linear mixtures. FastICA is a popular ICA algorithm due to its efficiency, but its performance strongly depends on a user-chosen nonlinear function matched to the source distribution.
When the source distribution is unknown, this function must be guessed at, and incorrect guesses can lead to significant drops in performance. We remove the need to guess by estimating a suitable function directly from the observed data. Our experiments show that the separation error stays close to the best fixed choice across synthetic mixtures comprising heavy-tailed or discrete sources while retaining a FastICA-like runtime.
\end{abstract}


\section{Introduction}

Let $X\in\mathbb{R}^m$ denote a random observation vector generated by a linear mixing model
\begin{equation}\label{eq:ica_model}
X=\mA S, \qquad \mA\in\mathbb{R}^{m\times m}\ \text{invertible},
\end{equation}
where $S\in\mathbb{R}^m$ is a random source vector whose components are mutually independent with unknown marginal distributions. We observe $N$ independent and identically distributed (i.i.d.) samples $\{\vx_n\}_{n=1}^N$ of $X$, arranged as the data matrix $\mX=[\vx_1,\ldots,\vx_N]\in\mathbb{R}^{m\times N}$. Likewise, let $\{\vs_n\}_{n=1}^N$ be the corresponding (unobserved) i.i.d. samples of $S$ and define $\mS=[\vs_1,\ldots,\vs_N]\in\mathbb{R}^{m\times N}$, so that in matrix form $\mX=\mA\mS$. The goal of ICA is to estimate a demixing matrix $\mW\in\mathbb{R}^{m\times m}$ where $\mY=\mW\mX$ is as close as possible to $\mathbf{S}$. $\mW$ is only identifiable up to permutation and scaling since for any permutation $\mathbf{P}$ and any invertible diagonal $\boldsymbol{\Lambda}$,
$\mX = (\mA\boldsymbol{\Lambda}^{-1}\mathbf{P}^{-1})(\mathbf{P}\boldsymbol{\Lambda}\mS)$.

FastICA~\cite{hyvarinen1999fast} is often used to solve the ICA problem because it is computationally efficient and numerically stable. However, its performance is strongly dependent on a user-chosen nonlinearity $g$ in its fixed-point update equation. Choosing $g$ implicitly assumes a family of source distributions; mismatch can slow convergence and increase the separation error, especially when the sources are skewed, heavy-tailed, or discrete. In many practical settings the marginal families are not known a priori, so selecting a single fixed $g$ becomes a tuning problem.

Prior approaches to reducing FastICA nonlinearity mismatch typically adapt $g$ from estimated source densities or related derivatives within ICA, or replace FastICA entirely with a separate characteristic-function objective. For instance, EFICA~\cite{koldovsky2006efficient} improves upon FastICA by adapting $g$ toward the score, i.e. the derivative of the log-density, to approach the Cramér-Rao bound, but this requires additional density estimation that can be sensitive and unstable beyond smooth, continuous marginals. Information-theoretic ICA methods (e.g., based on mutual information)~\cite{pham2004fast,learned2003ica} reduce dependence on a priori assumptions via nonparametric density estimation, but introduce bandwidth or binning choices and require repeated refits at each iteration, significantly increasing tuning complexity and runtime. Alternatively, characteristic-function-based (CF-based) ICA methods, such as CHFICA~\cite{eriksson2003characteristic}, offer consistency guarantees across diverse distributions~\cite{chen2005consistent}, but direct CF matching is computationally heavy compared to FastICA. While recent works have explored optimally weighted empirical characteristic function (ECF) estimation~\cite{starck2024ica} and CF-based diagnostics~\cite{kumar2024noisyica}, the computational burden of replacing the FastICA fixed-point update remains high.

To overcome the computational burden of direct CF matching and the sensitivity of iterative density estimation, we propose retaining the efficient FastICA fixed-point equation while replacing the manual selection of the nonlinearity $g$ with a function learned directly from the data. Motivated by recent advances in Fourier and ECF-based score learning outside of ICA~\cite{asokan2024fold}, we estimate a score function $\psi_Y(y)\triangleq \frac{d}{dy}\log f_Y(y)$ for the random variable $Y$ with density $f_Y$, using a low-frequency ECF aggregated across multiple one-dimensional Cramér-Wold projections. Note that the ``score'' here refers to the Stein score, the derivative of the log-density with respect to the \emph{observation}. By learning this univariate score-based nonlinearity once from the observed mixtures, we can tabulate $g(z)\approx -\psi(z)$ on a grid and access it via simple interpolation during FastICA. This preserves the per-iteration cost of the standard symmetric FastICA update.

\textbf{Contributions:} We propose a plug-in procedure that replaces FastICA’s user-chosen nonlinearity $g$ by a data-driven estimate learned once from the observed mixtures, allowing the FastICA fixed-point update and orthogonalisation to remain unchanged~\cite{hyvarinen1999fast}. Specifically:
\begin{enumerate}
    \item \emph{ECF-based score learning from projections:} we estimate a univariate score using projection-binned ECF (P-bECF) with equal-occupancy binning, subtractive dither, and safe-band sinc debiasing, then average across $R$ random projections.
    \item \emph{FastICA with learned nonlinearity:} we tabulate $g\approx -\psi$ and $g^\prime$ on a grid and substitute them into the FastICA fixed-point update equation.
    \item \emph{Empirical performance and efficiency:} on synthetic mixtures with continuous heavy-tailed or discrete marginals, we demonstrate that the learned nonlinearity achieves separation errors comparable to the best fixed-choice nonlinearity, while maintaining FastICA-like runtime.
\end{enumerate}

\textbf{Remark:} For noisy ICA, the same CF-domain correction used in~\cite{eriksson2003characteristic} can be applied when the noise CF is known and nonzero on the retained band, so we focus on the noise-free setting \eqref{eq:ica_model} for simplicity.

\section{Background}

\subsection{ICA with whitening}
We assume finite second moments, so that the covariance matrix $\Sigma_{\vx}$ exists, with $\Sigma_{\vx}=\mathbb{E}[\vx_n \vx_n^\top]$, and is nonsingular. Without loss of generality we work with zero-mean data by centering the observations, i.e. subtracting the sample mean $\bar \vx=\frac1N\sum_{n=1}^N \vx_n$ from each $\vx_n$; equivalently this corresponds to redefining the sources as
$\vs_n-\mathbb{E}[\vs_n]$.

A standard preprocessing step is whitening whereby one chooses any transform $\mV$ such that $\tilde \vx_n = \mV \vx_n$ satisfies $\mathbb{E}[\tilde \vx_n \tilde \vx_n^\top]=I$ (e.g., $\mV=\Sigma_{\vx}^{-1/2}$). Whitening is not unique since $\mQ \mV$ is also a whitening transform for any orthogonal $\mQ\in \mathbb{O}(m)$. Under ICA scale ambiguity, $\mA \vs_n = (\mA \mD)(\mD^{-1}\vs_n)$ for any invertible diagonal $\mD$, so we may adopt the normalisation $\mathbb{E}[\vs_n \vs_n^\top]=I$ (unit source variances) by absorbing component-wise
scalings into $\mA$. With this convention, there exists a choice of whitening transform (equivalently, a choice of $\mQ$)
for which the whitened mixing matrix $\tilde \mA \triangleq \mV \mA$ is orthogonal, i.e. $\tilde \mA\in \mathbb{O}(m)$. Consequently, after whitening the demixing problem reduces to estimating an orthogonal matrix $\mW\in \mathbb{O}(m)$ such that $\vy_n = \mW \tilde \vx_n$ has (approximately) independent components.

In what follows we assume data have been centered and whitened. To simplify notation, we overwrite
$\tilde \vx_n$ by $\vx_n$. Subsequently,
\begin{equation}
    \vy_n = \mW \vx_n,\qquad \mW\in \mathbb{O}(m) \subset \mathbb{R}^{m\times m},    
\end{equation}
and we estimate $\mW$ by minimising an objective function over $\mathbb{O}(m)$. Optimising over $\mathbb{O}(m)$ restricts the search to $m(m-1)/2$ degrees of freedom (rather than $m^2$) and keeps the demixing matrix well-conditioned in the sense that its singular values are fixed to one which, as well shall show in the sequel is desirable for FastICA.

\subsection{FastICA: symmetric fixed-point updates}

The one-dimensional projection of $\vx_n$ by $\vw_k$, the $k$th column of $\mW$, is $y_{k,n} = \vw_k^\top \vx_n$. We can find a $\mW$ such that $\mY=\mW\mX$ has independent rows by optimising a scalar objective built from these projections. Specifically
\begin{equation}\label{eq:fastica_obj}
    \max_{\mW\in \mathbb{O}(m)} \; \sum_{k=1}^m \mathbb{E}\!\left[G(\vw_k^\top \vx)\right]
    \quad \text{s.t.}\quad \mW\mW^\top = I.
\end{equation}
where $G:\mathbb{R}\to\mathbb{R}$ is a nonquadratic function and the expectation is taken with respect to $\vx$. The constrained maximisation \eqref{eq:fastica_obj} can be viewed as a tractable surrogate for minimising mutual information (MI): with whitening and an orthogonal $\mW$, reducing dependence is closely related to making each marginal $\vw_k^\top \vx$ as non-Gaussian as possible e.g. via negentropy approximations~\cite{hyvarinen1999fast}.

Introducing a symmetric matrix of Lagrange multipliers $\Lambda\in\mathbb{R}^{m\times m}$, the Lagrangian of \eqref{eq:fastica_obj} is
\begin{equation}\label{eq:lagrangian}
    \mathcal{L}(\mW,\Lambda)=\sum_{k=1}^m \mathbb{E}\!\left[G(\vw_k^\top \vx)\right]
    -\frac12 \operatorname{tr}\!\left(\Lambda(\mW\mW^\top-I)\right).
\end{equation}
Stationarity of $\mathcal{L}$ with respect to $\vw_k$ yields conditions of the form $\mathbb{E}[\vx\, g(\vw_k^\top \vx)] = \lambda_k \vw_k$. In practice, expectations are replaced by sample averages over $\{\vx_n\}_{n=1}^N$, and each row $\vw_k$ is updated by
\begin{equation}\label{eq:fasica_update}
    \vw_k \leftarrow \frac{1}{N}\sum_{n=1}^N \vx_n\, g(\vw_k^\top \vx_n)
    -\Big(\frac{1}{N}\sum_{n=1}^N g'(\vw_k^\top \vx_n)\Big) \vw_k,
\end{equation}
followed by symmetric orthogonalisation $\mW \leftarrow (\mW\mW^\top)^{-1/2}\mW$ to enforce $\mW\in \mathbb{O}(m)$ and prevent multiple rows from collapsing onto the same component\cite{hyvarinen1999fast}.

\paragraph{Choosing the nonlinearity} 
 
In MI based ICA~\cite{pham2004fast}, the (relative/natural) gradient of the MI criterion function involves the score functions $\psi_{Y_k}(y)=\partial_y \log p_{Y_k}(y)$ of the recovered components, so practical MI algorithms often rely on estimating these scores or replacing them with surrogate nonlinearities. In FastICA fixed-point iterations, the nonlinearity $g$ plays an analogous role; choosing $g \approx -\psi$ can therefore be interpreted as making the update consistent with a score-based descent direction for an information-theoretic / likelihood-related objective function (without requiring an explicit equivalence of fixed-point equations). 
In the large-sample regime, if the scores are identical to the true marginal score (or consistently estimated), this score-based choice yields the statistically optimal asymptotic accuracy, meaning that under regularity conditions the asymptotic error variances (the entries of the asymptotic covariance of $\sqrt{n}\,\mathrm{vec}(\widehat \mW-\mW)$ where $\mathrm{vec}$ is the vectorisation operator) are minimised and can attain the CRB~\cite{koldovsky2006efficient}. Conversely, a mismatch between $g$ and the true (unknown) scores leads to a loss relative to the CRB.

\subsection{Score function from the CF}\label{sec:score_function}

The CF of a random vector $X\in\mathbb{R}^m$ is
\begin{equation}\label{eq:cf_def}
\begin{aligned}
    \varphi_{X}(\mathbf{u}) \triangleq \mathbb{E}\!\left[e^{\,i\,\mathbf{u}^\top X}\right],\quad
    \mathbf{u}\in\mathbb{R}^m.
\end{aligned}
\end{equation}
with marginal CF $\varphi_{X_k}(u)\triangleq \varphi_X(0,\ldots,u,\ldots,0)$. Given i.i.d.\ samples $\{\vx_n\}_{n=1}^N$, the ECF is
\begin{equation}\label{eq:ecf_def}
\begin{aligned}
    \widehat{\varphi}_{X}(\mathbf{u})=\frac{1}{N}\sum_{n=1}^N e^{\,i\,\mathbf{u}^\top \mathbf{x}_n},
\end{aligned}
\end{equation}
and similarly for $\widehat{\varphi}_{X_k}$. For each fixed $\mathbf{u}$, $\widehat{\varphi}_X(\mathbf{u})$ is unbiased, and it is uniformly consistent on bounded frequency sets under standard conditions~\cite{ferreApplicationEmpiricalCharacteristic2004,ushakov2011selected}.
For $m=1$ Fourier inversion gives
\begin{equation}\label{eq:fourier_inversion_m}
    f_{X_k}(x)=\frac{1}{2\pi}\int_{\mathbb{R}} e^{-i\,u^\top x}\,\varphi_{X_k}(u)\,du.
\end{equation}
%e.g. when $\int_{\mathbb{R}}\left|u \phi_X(u)\right| d u<\infty$
Differentiating under the integral sign (permitted under mild regularity conditions), the univariate score function $\psi_{X_k}(x)\triangleq f^{\prime}_{X_k}(x)/f_{X_k}(x)$ admits the CF ratio representation
\begin{equation}
    \psi_{X_k}(x) \;=\; \frac{ -\frac{i}{2\pi}\int_{\mathbb{R}} u\,e^{-i\,u x}\,\varphi_{X_k}(u)\,du }
    { \frac{1}{2\pi}\int_{\mathbb{R}} e^{-i\,u x}\,\varphi_{X_k}(u)\,du }.
\end{equation}

In practice, we approximate these integrals using a truncated low-frequency grid (optionally tapered), which ensures the required regularity and yields a numerically stable score estimator.

\section{Learning the Score from Projection-Binned ECFs}

\subsection{Projection-binned ECF (P-bECF)}\label{sec:pbecf}
P-bECF replaces expensive multivariate CF estimation with
many simple 1-D probes along a set of directions $\{\mathbf{a}_r\}_{r=1}^R$,
and frequencies $\{u_\ell\}_{\ell=1}^L$, $R,L\in \mathbb{Z}_+$. We approximate 1-D CFs of projected data using binned histograms, then correct binning bias via subtractive dither and sinc debiasing.
\paragraph{Setup (whitened data)}
Assume $\mathbf{X}\in\mathbb{R}^{m\times N}$ has been centered and whitened, and let
$\varphi_{Z}(u; \mathbf{a})=\mathbb{E}[e^{iu\mathbf{a}^\top\mathbf{X}}]$ denote the
1-D CF of the projection $Z=\mathbf{a}^\top\mathbf{X}$ for a unit direction
$\mathbf{a}\in\mathbb{S}^{m-1}$. 

\paragraph{Low-frequency band}
In~\cite{ferreApplicationEmpiricalCharacteristic2004} the ECF variance is given by
$\operatorname{Var}\big(\widehat{\varphi}_{Z}(u; \mathbf{a})\big)\approx (1-|\varphi_{Z}(u;\mathbf{a})|^2)/N$ and is therefore small when $|\varphi_{Z}(u;\mathbf{a})|\approx 1$. For this reason, frequency selection is made on a low-$|u|$ band.

\paragraph{Dithered equal-occupancy binning with sinc debias}
Histogram discretisation schemes can be used to compute the ECF efficiently but data binning can introduce bias~\cite{silverman1986density,jones1983errors}. For each direction $\mathbf{a}_r$, $r\in\{1,\ldots,R\}$, we form $Z_r=\mathbf{a}_r^\top\mathbf{X}$ and quantise $Z_r$ to equal-occupancy bins of width $h$ after adding subtractive uniform dither $U[-h/2,h/2]$ to each sample. Next, $\{\hat p_{b,r}\}$, the binned empirical probabilities for direction $\mathbf{a}_r$ and bin centres $\{c_b\}$ are computed. Finally, the following ECF estimator for a single projection at frequency $u$ is obtained by combining the binned estimator of \eqref{eq:ecf_def} and $\operatorname{sinc}$ debiasing: 
\begin{equation}\label{eq:binned_ecf}
    \widehat{\varphi}^{\,\text{bin}}_{Z_r}(u)
    =\sum_b \hat p_{b,r}\, e^{iu c_b}\;/\;\operatorname{sinc}(u h/2).
\end{equation}
To avoid numerical instability near zeros we enforce a \emph{safe band} $|u|h\le c<\pi$ and floor the divisor by $\max\{\operatorname{sinc}(u h/2),\epsilon\}$ with small $\epsilon>0$.

\subsection{Projection-averaged score}
\paragraph{Band-limited score from CF ratios}
For any scalar $Z$ with CF $\varphi_Z$, the (univariate) score
$\psi_Z(z)=\frac{d}{dz}\log f_Z(z)$ admits the Fourier ratio representation (on our CF convention)
\begin{equation}
    \psi_Z(z)
    =
    \frac{-\frac{i}{2\pi}\int_{-T}^{T} u\,e^{-iuz}\,\varphi_Z(u)\,w(u)\,du}
    {\frac{1}{2\pi}\int_{-T}^{T} e^{-iuz}\,\varphi_Z(u)\,w(u)\,du},   
\end{equation}
with a smooth taper $w(u)$ (e.g., Gaussian) on $[-T,T]$. We estimate this by discrete weighted sums over $\{u_\ell\}_{\ell=1}^L$ using the dither-debiased P-bECF:
\begin{equation}\label{eq:score_est}
    \widehat{\psi}_{Z_r}(z)
    =
    \frac{-\,i\sum_{\ell=1}^{L} u_\ell\,e^{-iu_\ell z}\,
          w_\ell\,\widehat{\varphi}^{\,\text{bin}}_{Z_r}(u_\ell)}
         {\sum_{\ell=1}^{L} e^{-iu_\ell z}\,
          w_\ell\,\widehat{\varphi}^{\,\text{bin}}_{Z_r}(u_\ell) + \epsilon},
    \quad w_\ell\triangleq w(u_\ell).    
\end{equation}

\paragraph{Projection-averaging}
To reduce sensitivity to the projection, we average 1-D score estimates over $R$ directions:
\begin{equation}\label{eq:score_ave}
    \overline{\psi}(z)
    \;=\;
    \frac{1}{R}\sum_{r=1}^{R} \widehat{\psi}_{Z_r}(z).    
\end{equation}
Finally, approximate the non-linearity in the FastICA update equation \eqref{eq:fasica_update} by \eqref{eq:score_ave} i.e.
$g(z)\approx -\,\overline{\psi}(z)$.

% -------------------- Fig. 1 (GGD): span full width, 3-across --------------------
\begin{figure*}[!t]
  \centering
  \begin{subfigure}[t]{0.32\textwidth}
    \centering
    \includegraphics[width=\linewidth]{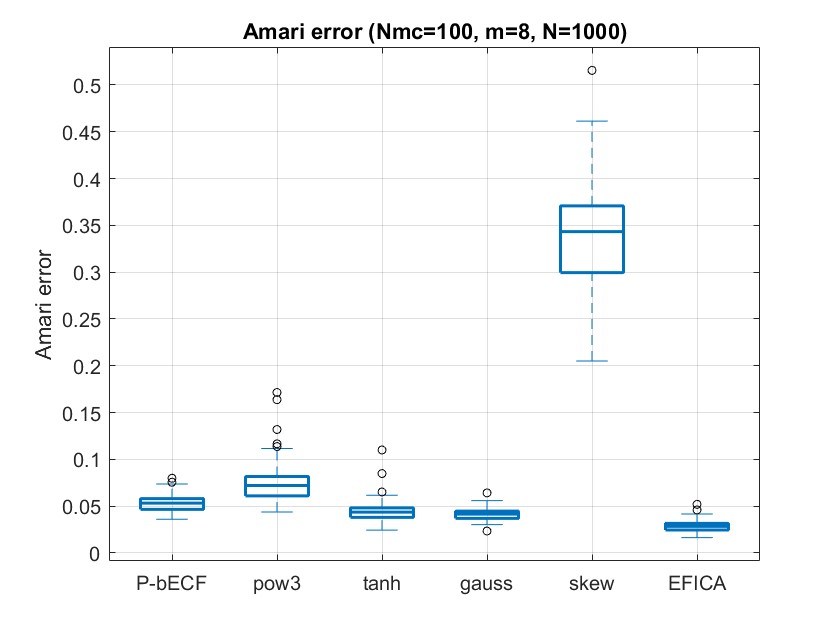}
    \subcaption{GGD ($\beta=1.6$) sources}\label{fig:ggd_8x8_amari}
  \end{subfigure}\hfill
  \begin{subfigure}[t]{0.32\textwidth}
    \centering
    \includegraphics[width=\linewidth]{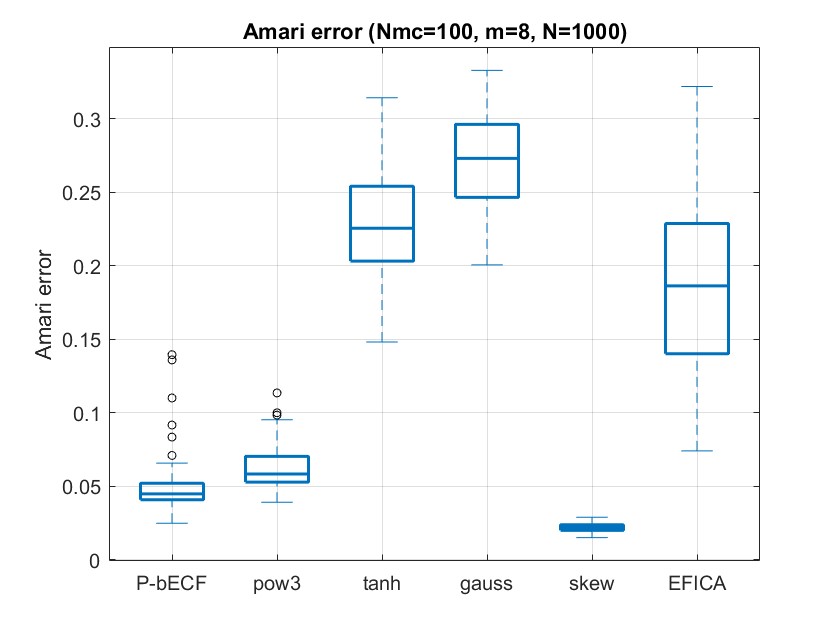}
    \subcaption{Poisson ($\lambda=0.5$) sources}\label{fig:poisson_8x8_amari}
  \end{subfigure}\hfill
  \begin{subfigure}[t]{0.32\textwidth}
    \centering
      \includegraphics[width=\columnwidth]{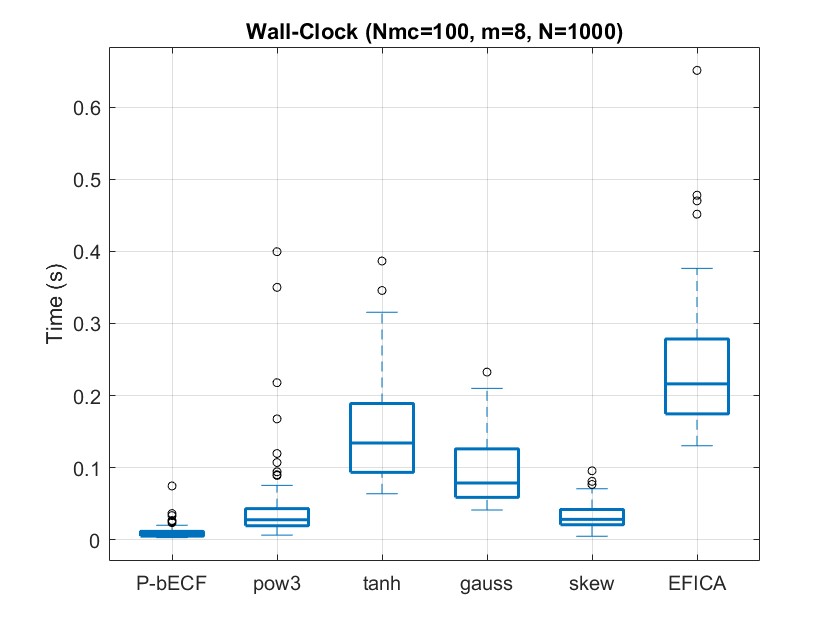}
      \caption{Wall-clock times.}
      \label{fig:wallclock_8x8}
  \end{subfigure}
  \caption{Performance over $N_{mc}=100$ Monte Carlo trials for $m=8$ sources; each panel summarises the distribution of trial-wise Amari error or runtime.}
  \label{fig:ggd_8x8_perf}
\end{figure*}

\section{Algorithm}\label{sec:algorithm}

\subsection{Pseudocode}

Pseudocode for our approach is provided in Algorithm \ref{alg:fastica_learned_score_full}. In all runs, $\mW_0$ is initialised by orthogonalising a Gaussian random matrix. Symmetric FastICA is run for at most $K_{\max}$ iterations, and stops early when the change between successive demixing matrices falls below convergence tolerance $\tau$. We use $K_{\max}=300$ and $\tau=10^{-6}$. The learned nonlinearity $g(y)\approx-\psi(y)$ is precomputed on a 1-D grid $\{z_j\}_{j=1}^J$ and evaluated during FastICA by interpolation; optionally one can re-estimate $g$ from intermediate outputs but this refinement is not considered.

\begin{algorithm}[t]
\caption{Symmetric FastICA with learned score via projection-binned, dither-debiased ECF}
\label{alg:fastica_learned_score_full}
\begin{algorithmic}[1]
\Require $\mX$, $R$, $\{u_\ell\}_{\ell=1}^L$, initial $\mW_0$, $\tau$ and $K_{\max}$.

\Statex \textbf{Score tabulation}
\State Sample directions $\{\va_r\}_{r=1}^R \subset \mathbb{S}^{m-1}$
\For{$r = 1,2,\dots,R$}
    \State $Z_r \gets \va_r^\top \mX$ %\Comment{1-D projections (length $n$)}
    \State Standardise $Z_r \gets (Z_r - \overline{Z}_r)/\mathrm{std}(Z_r)$
    \State Build equal-occupancy bins for $Z_r$
    \State Add subtractive dither; compute binned ECF $\tilde{\varphi}_r(u_\ell)$ on $\{u_\ell\}_{\ell=1}^L$
    \State Sinc-debias in the safe band using \eqref{eq:binned_ecf}
    \State Taper; compute $\widehat{\psi}_r(u_\ell)$ and $\widehat{\psi}'_r(u_\ell)$ using \eqref{eq:score_est}
\EndFor
\State Average across $R$ projections to obtain $\widehat{\psi}(u_\ell)$ and $\widehat{\psi}'(u_\ell)$ using \eqref{eq:score_ave}
\State Define interpolation handles $g(u)\gets \widehat{\psi}(u)$ and $g'(u)\gets \widehat{\psi}'(u)$

\Statex \textbf{FastICA (symmetric) with learned score}
\State $\mW \gets \mW_0$
\For{$k = 0,1,\dots,K_{\max}-1$}
    \State $\mY \gets \mW \mX$
    \State $G \gets g(\mY)$; \quad $G' \gets g'(\mY)$ %\Comment{elementwise}
    \State $\widetilde{\mW} \gets$ FastICA update using \eqref{eq:fasica_update}
    \State $\widetilde{\mW} \gets (\widetilde{\mW}\widetilde{\mW}^\top)^{-1/2}\,\widetilde{\mW}$ %\Comment{symmetric decorrelation}
    \If{$\|\widetilde{\mW} - \mW\|_F / \|\mW\|_F < \tau$}
        \State \textbf{break}
    \EndIf
    \State $\mW \gets \widetilde{\mW}$
\EndFor
\State \Return $\mW,\, g,\, g'$

\end{algorithmic}
\end{algorithm}

\subsection{Inputs and defaults}\label{sec:defaults}

Our score-learning step has four main discretisation and averaging parameters: the number of projection directions $R$, the histogram resolution (number of bins $B$ and bin width $h$) used to approximate each projection $Z_r=\va_r^\top \mX$, the number $L$ of low-frequency CF samples retained for the score estimate, and the tabulation grid size $J$ used to interpolate the learned nonlinearity $g\approx-\psi$ during FastICA. These parameters trade statistical error
(variance and conditioning), numerical approximation error, and computation.

\paragraph{Number of projections $R$}
Identifiability of a distribution from projections requires a totality of all directions (not just $R$)~\cite{heppes1956determination}. Since the projection-average variance from \eqref{eq:score_ave} decreases approximately as $1/R$, we choose $R$ as a small constant that stabilises the learned $g$ across random draws of $\{a_r\}$. For small dimensions ($m\le 8$), we find that $R\in\{8,\ldots,16\}$ is sufficient (we use default $R=12$); very large $R$ can average away informative structure in discrete distributions.

\paragraph{Histogram resolution ($B$ bins, width $h$)}
For each standardised projection $Z_r$, we approximate its distribution by binning $N$ samples into $B$ bins (width $h$ ). The width $h$ trades discretisation error against numerical conditioning of the sinc debiasing factor $1 / \operatorname{sinc}(u h / 2)$; we enforce a safe band with $c=0.3$ and stabilise the divisor with a Tikhonov floor $\delta \approx 10^{-3}$ to avoid amplification near sinc zeros. Increasing $B$ decreases $h$ (lower bias, wider usable low-frequency band) but increases histogram variance when $N / B$ is small and raises the FFT cost $O(B \log B)$. We choose $B$ to keep average occupancy moderate (e.g. $N / B \gtrsim 30$ ) while ensuring the safe band covers the frequencies used by the score estimator. In our experiments, $B\in\{64,\dots,256\}$ works well; we use $B=128$ by default and reduce $B$ for smaller $N$.

\paragraph{Number of retained CF frequencies $L$}
Although the FFT of the histogram provides CF samples on an $O(B)$ frequency grid, we retain only $L\ll B$ low-frequency points $\{u_\ell\}_{\ell=1}^L$ for computing the score via Fourier inversion. The estimator~\eqref{eq:score_est} is most stable when the retained frequencies lie near the origin where the ECF has lowest variance for small $|u|$ ($|\varphi_{Z_r}(u)|\approx 1$). Including large $|u|$ increases Monte-Carlo variability and can destabilise the ratio indirectly by degrading the estimate of $f_{Z_r}(z)$ in regions where it is small. Accordingly, we select $\{u_\ell\}$ within the sinc-debias safe band $|u|h\le c$ and use a small $L$. Errors on the retained frequency grid act through the CF-ratio estimator as a bias-variance trade-off: too many or too-large frequencies increase variance and can destabilise the denominator, whereas too narrow a band oversmooths the learned score. Our pilot runs suggest that a default of $L=5$ captures the low-frequency content relevant for score tabulation.

\paragraph{Tabulation grid size $J$}
Whereas $R$, $B$, and $L$ primarily affect statistical error through the P-bECF estimator, $J$ controls the numerical approximation error of the tabulated FastICA nonlinearity $g$ and its derivative $g^\prime$.
We tabulate $g$ on a uniform grid over a bounded interval $[-z_{\max}, z_{\max}]$ with spacing $\Delta z = 2z_{\max}/(J-1)$, and evaluate $g$ during iterations by linear interpolation. When the learned $g$ is smooth, the interpolation error decays rapidly with $\Delta z$ (for linear interpolation, typically on the order of $\Delta z^2$ for twice-differentiable targets). We set $z_{\max}$ using robust quantiles of the projected samples (e.g. $z_{\max} = \mathrm{quantile}(|Z|, 0.995)$) and choose the smallest $J$ beyond which there is negligible improvement in separation metric. From pilot runs, we chose $J \approx 64$, beyond which performance changes negligibly.

\paragraph{Computational profile}

Each projection $Z_r=\mathbf{a}_r^\top \mX$ costs $O(mN)$, building the (equal-occupancy) histogram costs $O(N)$ (or $O(N\log N)$ if bin edges are obtained by sorting) and computing the binned ECF via an FFT of length $B$ costs $\mathcal{F}(B)=O(B\log B)$. Evaluating the score estimate in~\eqref{eq:score_est} on the $J$-point $z$-grid from $L$ frequencies costs $O(JL)$. Thus, the total tabulation cost is $O\!\left(R\,(mN + \mathcal{F}(B) + JL)\right)$. Thereafter, symmetric FastICA iterations have the standard cost dominated by matrix–data products ($O(m^2N)$ per iteration for $\mathbf{Y}=\mathbf{W}\mathbf{X}$ and $g(\mathbf{Y})\mathbf{X}^\top$), since $g$ and $g^\prime$ are obtained by table lookup and
interpolation.

In pilot runs, we verify that the learned $g$ and final separation metrics are stable to doubling each of $(R,B,L,J)$ one at a time; if not, we increase the parameter that induces the largest change. The safe-band restriction, tapering, and this pilot stability check are intended to keep finite-band approximation error small in practice.

\section{Experiments}\label{sec:experiments}

The goal of these experiments is to show that our method per Algorithm~\ref{alg:fastica_learned_score_full} has a separation performance that is at least as good as the FastICA variant that has the best user-chosen $g$ across different marginal families, and for the same computation cost. FastICA variants include $g\in \{\tanh,\operatorname{pow3},\operatorname{skew},\operatorname{gauss}\}$\cite{hyvarinen1999fast}, and we include EFICA~\cite{koldovsky2006efficient} as a near-CRB performance reference. Using Matlab, we perform two experiments each of $N_{mc}=100$ Monte Carlo trials. In the first experiment we use mixture of sources distributed as generalised Gaussian (GGD, $\beta=1.6$), and Poisson $(\lambda=0.5)$ for the second. We use $m=8$ sources and $N=1000$ observations for both experiments. Using a single marginal family across all $m$ sources in this manner is typical for ICA since it isolates the effect of nonlinearity mismatch without conflating it with per-component heterogeneity. We report metrics on two axes:
\begin{enumerate}
    \item We report Amari's separation error $E_{\mathrm A}$~\cite{amari1995new}, defined in (14). It quantifies how far the gain matrix $\mathbf{P}=\left(p_{i j}\right)=\mathbf{W} \mathbf{A}$ is from an ideal scaled permutation (perfect demixing), and $E_{\mathrm A}=0$ indicates perfect separation.
\begin{equation}
   E_{\mathrm A}=\sum_{i=1}^n\left(\frac{\sum_{j=1}^n\left|p_{i j}\right|}{\max _k\left|p_{i k}\right|}-1\right)+\sum_{j=1}^n\left(\frac{\sum_{i=1}^n\left|p_{i j}\right|}{\max _k\left|p_{k j}\right|}-1\right). 
\end{equation}
    \item Wall-clock time, measured under single-threaded Matlab execution on an Intel Core i7 CPU with 32 GB RAM.
\end{enumerate}

Fig.~\ref{fig:ggd_8x8_amari} shows the Amari error of FastICA with P-bECF score learning (labelled ``P-bECF'') on symmetric continuous mixtures. The FastICA variants $\{\tanh,\mathrm{pow3},\mathrm{gauss}\}$ have low error due to their sensitive to kurtosis. Our method is comparable which we attribute to the low variance in score estimation arising from point selection in a low-frequency band. At the size of this data record, EFICA is also competitive on well-behaved continuous sources as expected from its score-adaptive design.

Separation performance of discrete Poisson mixtures are shown in Fig.~\ref{fig:poisson_8x8_amari}. FastICA with the skew-sensitive nonlinearity $\mathrm{skew}$ now demonstrates near-CRB efficiency. P-bECF remains competitive, whereas the performance of $\tanh$ and $\mathrm{gauss}$ has deteriorated.

Wall-clock times across trials are conveyed as box plots in Fig.~\ref{fig:wallclock_8x8} confirming P-bECF achieves FastICA runtime and needs fewer iterations to converge. Overhead due to tabulation cost is minimal because (i) few ($L$) points are required to characterise the marginal distributions of the projected data and (ii) score estimation parameters $R,J,B$ have been kept to a minimum through pilot selection.

\section{Conclusion}

We presented a FastICA variant that eliminates manual nonlinearity selection by learning a score-motivated nonlinearity once from the mixtures using P-bECF. The learned $g$ is computed from a small set of low-frequency CF samples, numerically stabilised by projection averaging, subtractive dither, and safe-band sinc debiasing, then tabulated and reused via interpolation so that the per-iteration cost matches conventional symmetric FastICA. Experiments on both continuous heavy-tailed and discrete source mixtures demonstrate the intended benefit: when the source family is unknown and the fixed nonlinearities are prone to mismatch, the learned nonlinearity remains competitive across regimes while preserving FastICA runtime.

%%%%%%%%%%%%%%%%%%%%%%%%%%%
\clearpage
\bibliographystyle{IEEEtran}
\bibliography{./refs}

\end{document}